# Extraordinary Magnetooptics in Plasmonic Crystals

V.I. Belotelov[1,2], I.A. Akimov[3,4], M. Pohl[3], V.A. Kotov[1,5], S. Kasture[6], A.S. Vengurlekar[6], A.V. Gopal[6], D. Yakovlev[4], A.K. Zvezdin[1], M. Bayer[3]

1) A.M. Prokhorov General Physics Institute, Russian Academy of Sciences, 119991 Moscow, Russia
2) M.V. Lomonosov Moscow State University, 119992 Moscow, Russia
3) TU Dortmund University, D-44221 Dortmund, Germany
4) A.F. Ioffe Physical-Technical Institute, Russian Academy of Sciences, 194021 St. Petersburg, Russia
5) V.A. Kotelnikov Institute of Radio Engineering and Electronics, Russian Academy of Sciences, 125009 Moscow, Russia
6) Tata Institute of Fundamental Research, Mumbai, 400005, India
Email: Belotelov@physics.msu.ru

**Abstract**

Plasmonics has been attracting considerable interest as it allows localization of light at nanoscale dimensions. A breakthrough in integrated nanophotonics can be obtained by fabricating plasmonic functional materials. Such systems may show a rich variety of novel phenomena and also have huge application potential. In particular magnetooptical materials are appealing as they may provide ultrafast control of laser light and surface plasmons via an external magnetic field. Here we demonstrate a new magnetooptical material: a one-dimensional plasmonic crystal formed by a periodically perforated noble metal film on top of a ferromagnetic dielectric film. It provides giant Faraday and Kerr effects as proved by the observation of enhancement of the transverse Kerr effect near Ebbesen's extraordinary transmission peaks by three orders of magnitude. Surface plasmon polaritons play a decisive role in this enhancement, as the Kerr effect depends sensitively on their properties. The plasmonic crystal can be operated in transmission, so that it may be implemented in devices for telecommunication, plasmonic circuitry, magnetic field sensing and all-optical magnetic data storage.

**Introduction**

Plasmonics merges photonics and electronics at the nanoscale [1-5] and may offer breakthrough solutions for fully integrated photonics, allowing one to squeeze light into nanodevices. Widely spanned suggestions for plasmonic applications concern telecommunication, plasmonic circuitry, nanolithography, solar cells etc. [1-5]. Surface plasmon polaritons (SPPs) – coupled oscillations of the electromagnetic field and the electron plasma in a metal – play a decisive role in plasmonics. A prominent feature of SPPs is that they concentrate electromagnetic energy in a nanoscale volume near a metal/dielectric interface and thus increase light – matter interaction, leading to an enhancement of nonlinear effects and Raman scattering [6,7]. The huge potential of plasmonics has still not been fully explored so far, and the discovery of many interesting phenomena can be expected. In this respect, functional materials that change their properties due to an external stimulus are very prospective.

Here we consider a plasmonic structure that incorporates a magnetic medium making the SPPs and therefore also light sensitive and controllable by an external magnetic field. We demonstrate that such a tailored plasmonic nanostructure leads to a huge enhancement of magnetooptical effects. The problem of plasmonic materials considered so far has been the smallness of magneto-optical effects, excluding basically their application. This problem has been solved now as the increase of magneto-optical effects can reach 3 orders of magnitude at room temperature.

Examples of magnetooptical effects are the Faraday effect and the family of Kerr effects. They provide rotation of the polarisation of transmitted light by the Faraday effect and of reflected light by the polar and longitudinal Kerr effects [8]. As such, they are highly interesting for modulation of light beam intensities. If the applied magnetic field is perpendicular to the light incidence plane a magnetisation induced change of the reflected light intensity is observed which is referred to as the transverse magnetooptical Kerr effect (TMOKE) [8].

The magnetooptical effects in smooth single ferromagnetic films are usually not sufficiently large. It is this smallness which has been preventing an invasion of magnetooptics into applied nanophotonics. Therefore it is essential to seek for a strategy to enhance magneto-optical effects, for which there are several ways. While the potential of methods relying on material synthesis is almost exhausted, nanostructuring is very prospective for tailoring the optical properties of materials [9]. This approach reflects a new paradigm of modern optics in which optical properties are mainly determined by Fano-type geometrical resonances rather than by electronic ones. An example of a fruitful magnetooptical implementation of this approach is a magnetophotonic crystal providing a considerable enhancement of the Faraday effect [10-12].

Plasmonics provides another approach to boost magnetooptical effects. Early papers addressed SPPs propagating along the smooth surface of a ferromagnetic film [13-15] or along a smooth semiconductor surface in a transverse external magnetic field [16]. In that case, the magnetic field modifies the SPP wave vector but leaves its transverse magnetic polarization (TM) unchanged. An SPP assisted increase of the TMOKE at a ferromagnetic metal surface was reported [17-20] as well as an enhancement of the TMOKE in bimetallic systems consisting of a noble and a ferromagnetic metal [21-23]. Nanostructured plasmonic systems were investigated in [24-29], where an increase of the TMOKE was reported for one-dimensional Co and Fe gratings [24,25], for noble metals in a high external magnetic field [26], and for noble-metal/ferromagnetic-metal multilayers with periodic perforations or protrusions [27,28].

The main disadvantage of most of the aforementioned approaches is that the optical losses associated with the presence of a ferromagnetic metal are relatively high. This limits exploiting fully the potential gain of the combined concepts of nanostructuring and plasmonics in magneto-optics. If the ferromagnetic metals were avoided as in the case of pure semiconductor or noble metal systems, huge external magnetic fields exceeding several Tesla would be necessary to make the TMOKE at least comparable with the effect in ferromagnets.

In this paper we pursue the concepts of nanostructuring and plasmonics by demonstrating a new type of nanophotonic material that is a plasmonic crystal consisting of two thin films, a smooth ferromagnetic dielectric below a noble metal perforated with subwavelength slits (Fig.1). The proposed material therefore exhibits gyrotropic, plasmonic and nanopattern features. Because of its constituents it has a large specific Faraday rotation due to the ferromagnetic dielectric (the bismuth rare-earth iron garnet in our case) and record small optical losses for wavelengths above 650 nm (due to the nanostructured noble metal). Our previous theoretical investigations showed that in these structures the magnetooptical Faraday and Kerr effects can be resonantly increased, in particular near the Rayleigh-Wood resonances [30,31]. Here these predictions were assessed by fabricating the proposed structure and demonstrating outstandingly enhanced TMOKE (by about $10^3$ times) in transmission. Our development therefore represents a crucial step towards advanced magnetooptics.

**Surface magnetoplasmons, Wood anomalies and extraordinary magnetooptics**

TMOKE is usually characterized by the relative change δ of reflected light intensity $R(\mathbf{M})$ when a medium's magnetization $\mathbf{M}$ is reversed:

$$\delta = \left(R(\mathbf{M}) - R(-\mathbf{M})\right)/R(0). \tag{1}$$

This change originates from the magnetic field induced change of boundary conditions at the magnetic layer surface, and attains maximum values for oblique incidence of p-polarized light (electric field vector $\mathbf{E}$ parallel to the incidence plane) while it almost vanishes for s-polarization [8]. Usually, for smooth ferromagnetic metals δ is on the order of $10^{-3}$ setting strong limits to its applicability [8,32]. The TMOKE's counterpart in transmission may also occur, for which the necessary condition is a difference between the opposite magnetic-film boundaries. However, it is difficult to be observed because of the small transmission through ferromagnetic metals in addition to its small magnitude [33]. We in what follows concentrate on the observation of the TMOKE in the magnetoplasmonic crystal in transmission.

In absence of an external magnetic field the magnetization $\mathbf{M}$ of the ferromagnetic film used in our experiments (see Suplementary-2) is perpendicular to its surface. It can be reoriented parallel (Fig.1 a) or perpendicular (Fig.1 b) to the slits in the noble metal by application of an external magnetic field. The incident light is either p-polarised or s-polarised (electric field **E** perpendicular to the incidence plane). The incidence plane is perpendicular to **M**.

Using a linear approximation for the dependence of the permittivity tensor describing the ferromagnetic film on the gyration (the magnetooptical parameter) this tensor is given by:

$$\hat{\varepsilon}_m = \begin{pmatrix} \varepsilon_2 & 0 & ig_y \\ 0 & \varepsilon_2 & -ig_x \\ -ig_y & ig_x & \varepsilon_2 \end{pmatrix}, \qquad (2)$$

where $g_x$ and $g_y$ are the components of the gyration vector $\mathbf{g} = a\mathbf{M}$ [8]. The magnetic tensor $\mu$ is taken to be unity, since the magnetic dipole response at optical frequencies is very weak [34]. The gold metallic layer is characterized by the dielectric function $\varepsilon_1$ [35].

There are several vectors that characterize the optical and magnetooptical response of the system. These are the incident light wave vector **k**, the polarization vector **P**, and the lattice vector **D** (see Fig. 1a). Together they determine the necessary condition for the excitation of SPPs in the slit grating system: $(\mathbf{P} \cdot \mathbf{D}) \neq 0$ and they also determine the SPPs quasimomentum $\mathbf{k}_{SPP}$:

$$\mathbf{k} = \mathbf{k}_{SPP} + (2\pi/D)m\mathbf{e}_x, \qquad (3)$$

where $m$ is an integer and $\mathbf{e}_x$ is a unit vector along the x-axis. Thus, in the conventional configuration (Fig. 1a) SPPs can be generated by p-polarised light, while for the second configuration (Fig. 1b) it is the s-polarisation which can excite SPPs.

The other two important vectors are the magnetization **M**, and a vector **N** normal to the metal-dielectric interface. The necessary condition for the occurrence of TMOKE is $[\mathbf{k} \times \mathbf{N}] \neq 0$. The cross product $[\mathbf{M} \times \mathbf{N}]$ is also very important. It is nonzero near the surface of the magnetized film. The magnetic field breaks the symmetry with respect to time reversal, while the interface (and the **N** vector normal to it) breaks the space inversion. Interestingly, space–time symmetry breaking is characteristic of media with a toroidal moment $\tau$ which has transformation properties similar to the ones of $[\mathbf{M} \times \mathbf{N}]$ [36]. Consequently, the problem of SPP propagation along the interface of a transversely magnetized medium is similar to that of electromagnetic wave propagation in a bulk medium with a toroidal moment parallel to this direction. In electrodynamics, a toroidal moment is known to give rise to optical nonreciprocity as manifested by a difference between the wave vectors for waves propagating forward and backward with respect to the toroidal moment [37]. An expression for optical nonreciprocity can be derived for

SPPs by solving the Maxwell equations for metallic and magnetic layers with appropriate boundary conditions [23,31]. For a smooth metal/magnetic-dielectric interface it is given by

$$k_{spp} = \frac{\omega}{c}\sqrt{\frac{\varepsilon_1\varepsilon_2}{\varepsilon_1+\varepsilon_2}}(1+\alpha g_y), \qquad (4)$$

where $\omega$ is the incident light frequency, $c$ is the light velocity in vacuum, and $\alpha = (-\varepsilon_1\varepsilon_2)^{-1/2}(1-\varepsilon_2^2/\varepsilon_1^2)^{-1}$.

The nonreciprocity effect is a prominent inherent feature of the SPP assisted TMOKE, as can clearly be seen for a smooth metal/dielectric interface. If one works far from the plasmonic resonance one usually observes a monotonic featureless reflection spectrum and the TMOKE signal is quite small even for metal ferromagnets. On the contrary, at the SPP resonance a pronounced dip appears in the reflection spectrum. Since the SPP wave vector differs for opposite magnetizations according to (4), the reflection dip shifts with magnetic field to smaller or higher frequencies and the TMOKE signal becomes enhanced by about an order of magnitude. Actually, in the frequency range of SPP generation, the TMOKE parameter $\delta$ is approximately equal to a product given by the frequency derivative of the reflection spectrum and the magnetic induced frequency shift. The steeper the resonance is, the larger is the TMOKE. Thus, TMOKE enhancement through SPPs occurs even for a smooth interface. However, there are drawbacks in this approach. Firstly, to excite SPPs at the interface between a metal and a magnetic film, the refractive index of the prism (in Kretschmann geometry) must be larger than the refractive index of the magnetic film, which is rather difficult to fulfill. Secondly, sputtering of the opaque metal layer onto the dielectric reduces the transmission to zero preventing operation in transmission. That is why a transfer of the hybrid approach to systems with perforated metal may be worth to be considered.

Ebbesen et al. [38] demonstrated the phenomenon of extraordinary optical transmission in perforated metal films where pronounced peaks appear in the transmission spectrum so that the structure is much more transparent than a smooth gold film. This phenomenon is present even for very narrow slits, and allows one to fabricate a perforated sample with large transmission resonances and the possibility to excite SPPs almost undisturbed by the metal film discontinuities. With such a structure we expect that it may be possible to observe TMOKE in transmission, which is of prime importance for applications.

The exact analytical description of such a system with periodic nanostructuring is extremely difficult. Nevertheless, a semiquantitative analysis is feasible. The SPPs in the periodic system are characterized by their quasimomenta $k_{spp}$ in the first Brillouin zone. It can be shown [39] that

the SPP excitation frequency for the magnetized sample is shifted from the non-magnetized one by

$$\Delta\omega_{spp}(g) = \xi \frac{k_{spp} g_y}{\omega_0(k_{spp})\varepsilon_2^2},  \qquad (5)$$

where $\xi$ is a coefficient depending on the electromagnetic field distribution at the interface, and $\omega_0(k_{spp})$ is the eigenfrequency of the SPP in the nonmagnetic case. Nonreciprocity is also present in the perforated magnetoplasmonic structure. It is governed not only by the gyration vector and the dielectric constants of the heterostructure constituents, but also by the electromagnetic field distribution at the interface, i.e., it depends on the grating parameters. Moreover, since the TMOKE in this case resembles the frequency derivative of the resonances in the reflection or transmission spectrum, it also depends strongly on the shape of these spectra.

It was shown that the Wood anomalies [40], i.e. the features in transmission and reflection spectra related to SPP excitation in metal gratings, are Fano resonances having a characteristic asymmetric profile with a maximum followed by a minimum (or vice versa) [41]. The origin of the Fano shape is interference between resonant processes in the excitation of eigenmodes of the structure, such as quasi-waveguide modes, SPPs, slit modes, etc. plus a nonresonant contribution from the radiation directly scattered by the grating without excitation of eigenmodes. If the nonresonant process is negligible, then the frequency dependence has the standard symmetric shape of a Lorentz curve.

Since SPPs can propagate along both surfaces of the perforated metal film, two types of Fano resonances will be observed. However, it is only the bottom metal surface which is adjacent to the ferromagnetic layer that contributes essentially to the TMOKE. Consequently, we can assume that the TMOKE enhancement around the two resonances will be significantly different, exhibiting obviously a much larger enhancement factor for the SPPs on the bottom interface. Furthermore, contrary to uniform films, the optical properties of perforated metals are governed not only by SPPs but by other eigenmodes and anomalies as well [42,43]. Indeed, dips and peaks in the transmission/reflection spectra might also be due to Rayleigh anomalies or Fabry-Perot resonances. Such diversity of optical phenomena usually leads to considerable difficulties in their interpretation. Interestingly, TMOKE observation can reveal the difference between them: since Rayleigh anomalies (sharp maxima in the reflection spectra of the metallic gratings) are related to electromagnetic field singularities when one of the diffracted orders becomes tangent to the grating surface and are determined by the grating period, any influence of the magnetization on them is impossible. Also the anomalies caused by Fabry-Perot resonances inside the slits should not be very sensitive to the magnetization as they are mainly determined by slit depth and width [42].

So far we analyzed only the first configuration for incidence of p-polarised light. If in the same configuration the light is s-polarised, then no SPPs can be excited. It follows from the form of the tensor $\hat{\varepsilon}_m$ that if $\mu = 1$ then the TMOKE should also vanish. This corresponds to the "non-plasmonic" and "non-TMOKE" case (see Table in Fig. 5). If s-polarized light is incident along the slits (Fig. 1b) then SPPs are excited. However, no TMOKE is possible since the light is polarised perpendicularly to the incidence plane. So here we are dealing with the "plasmonic" and "non-TMOKE" case. Finally, if in this configuration the light is p-polarised, SPPs are not excited but TMOKE should be present. Thus we can label this situation as "non-plasmonic" and "TMOKE", so that the TMOKE cannot be enhanced by SPPs.

**Giant TMOKE in transmission mode**

A more detailed analysis of the nanostructured system can be performed on the basis of numerical modeling (see Supplementary 1). Preliminary numerical modeling allowed us to design the sample, i.e. determine the gold grating period, gold thickness, and slit width, and to adjust the main SPP resonances to the wavelength range of 650 – 850 nm, which is most suitable for magnetooptical experiments on bismuth iron garnets. The magnetooptical figure of merit given by the ratio of the specific Faraday rotation to the absorption is highest around 750 nm.

We start from the investigation of the first experimental configuration (Fig. 1a), corresponding to the "plasmonic" and "TMOKE" case. Results of the experimentally measured zero-order transmission for this case are shown in Fig. 2 a (see Supplementary 3). Comparison with the calculated band structure (Fig. 2c) allows us to attribute the pronounced Fano resonance (1) in Fig. 2a to the Wood's anomaly of the 2$^{nd}$ band SPP at the air/gold interface, while the Fano resonances (2) and (3) are related to the 2$^{nd}$ and 3$^{rd}$ band SPP at the gold/magnetic-film interface. Finally, the prominent transmission peak (4) is attributed to the collective Fabry-Perot cavity mode inside the slits.

The experimentally measured TMOKE parameter δ is defined in accordance to (1) but with the reflection coefficient substituted by the transmission coefficient (Fig. 2b). To ensure that the sample magnetization is oriented almost completely in the plane, a relatively large external magnetic field of 2000 Oe was applied. Outside of the resonances the absolute value of δ is very small. Actually in this case, δ cannot be measured experimentally, which means that it is below $10^{-3}$. On this background, pronounced positive (red) and negative (blue) peaks are observed at which δ reaches up to $1.5 \cdot 10^{-2}$ demonstrating a TMOKE increase by at least one order of magnitude. Electromagnetic modeling for the nonresonant case gives δ=$8 \cdot 10^{-4}$ implying an enhancement factor of about 20. Compared to the uncovered bare iron garnet film, the enhancement factor is much larger, about $10^3$. From these results we can claim that a giant TMOKE has been observed in transmission. It should be noted that here we use a magnetic film with a relatively small concentration of bismuth. For iron garnets with a composition $Bi_3Fe_5O_{12}$ the specific Faraday rotation is about 6° at λ=630 nm, 13 times larger than for our sample [8]. Since the δ-value is proportional to the gyration (and to the specific Faraday rotation), δ should be also 13 times larger than the one observed in our experiments, i.e. δ may exceed 0.2.

The regions of enhanced TMOKE clearly correspond to the regions of SPP excitation at the gold/ferromagnet interface (compare Figs. 2a and 2b). No notable TMOKE increase is observed for other resonant regions, in agreement with the discussion above. This highlights the TMOKE's sensitivity to the excitation of different eigenmodes.

**TMOKE as a new tool for SPP probing**

A close-up of the TMOKE spectral shape is shown in Fig. 3. For normal incidence the TMOKE is zero because of the symmetry of the incident light with respect to the structure (Fig. 3a). The magnetization induced shift of the SPP resonance frequency vanishes (see (5)) and there is degeneracy of the SPPs traveling forward and backward. When **k** is not normal to the surface, the symmetry is broken and the degeneracy is lifted. SPP modes propagating in opposite directions are excited at slightly different frequencies and TMOKE appears. δ reaches almost $10^{-2}$ even if the incidence angle is as small as θ=0.8° (Fig. 3a), in resonance with the Fano resonance for SPPs on the gold/magnetic-dielectric interface in the vicinity of the Γ point of the first Brillouin zone. Going from normal to slightly oblique incidence, the transmission spectrum, however, does not change notably, in contrast to the TMOKE spectrum. This demonstrates sensitivity of the TMOKE to the SPP modes in the structure. No measurable TMOKE signal is observed around the other features in the transmission spectrum, the peak at λ=675 nm and the dip at λ=675 nm.

When the incidence angle becomes larger (e.g. θ=5°, Fig. 3b), the eigenfrequencies of the two SPPs propagating in opposite directions differ significantly which gives rise to two Fano resonances in transmission. TMOKE accompanies both resonances, but with opposite signs of δ, reflecting the fact that these resonances are due to SPPs propagating in opposite directions with respect to the cross product $[\mathbf{M} \times \mathbf{N}]$. This observation unravels another prominent feature of TMOKE in magnetoplasmonic structures: through the sign of δ one can distinguish between resonances caused by SPPs propagating in opposite directions.

For larger angles of incidence the two Fano resonances split further apart and the TMOKE signal does not change much (Fig. 3c). It should be noted that similar measurements for negative angles of incidence give the same value of δ but with opposite sign. This demonstrates that the observed effect is odd in magnetization and not an experimental artifact (Fig. 3d). Above some incidence angle (e.g. at θ=15°) the SPP resonance is barely detectable but the TMOKE still indicates its frequency position (see peak (i) of δ in Fig.4a). Thus TMOKE allows measurement of the energy spectrum of the SPP eigenmodes at a metal/ferromagnetic interface.

The magnetic field dependence of δ is shown in Fig. 4b for the two main TMOKE peaks (i) and (ii) at θ=15°. At small magnetic fields the value of δ grows linearly with magnetic field. In accordance to (5) this indicates that the in-plane component of the sample magnetization also has a linear dependence on magnetic field. Saturation takes place at a magnetic field strength of about 1600 Oe which is in nice agreement with the predicted micromagnetic properties of the magnetic film, namely with the value of the effective uniaxial magnetic anisotropy field (see Supplementary 2). However, δ reaches relatively high values even for smaller fields. For example it is $5 \cdot 10^{-3}$ at 300 Oe field which by far exceeds the noise level and is easily measurable.

Let us now consider the other possible experimental configurations. In accordance to our analysis no TMOKE and no SPP excitation are possible for s-polarized incident light and **M** parallel to the slits (Fig. 1a). This is confirmed by the absence of the extraordinary peaks in Fig. 5 (curve 1). The incidence of s-polarised light parallel to the slits direction (Fig. 1b) gives rather pronounced transmission peaks proving excitation of SPPs and cavity eigenmodes (Fig. 5, curve 2). As expected no TMOKE is present though, which is approved by numerical calculations. Finally, the "non-plasmonic" but "TMOKE" case is considered. For p-polarised light the absence of SPPs is clearly indicated by the low transmission (Fig. 5 curve 3). TMOKE signal cannot be measured experimentally. This is not surprising since the numerical modeling predicts δ to be no more than $10^{-4}$ which is below our experimental setup resolution. This case emphasizes the role of SPPs for the TMOKE enhancement: If SPPs are not excited, δ is reduced by two orders of magnitude.

**Conclusion**

We proposed and fabricated a new magnetooptical material that is a plasmonic crystal consisting of a smooth epitaxial magnetic film covered by a thin gold layer perforated with a periodic subwavelength slit array. Such a structure has three important features, namely, (i) it is periodically nanostructured, (ii) it is plasmonic, and (iii) it is magnetic. The main aim of our study was to identify how the first two features affect the magnetooptical properties of the structure and in particular whether they can be used to enhance the magnetooptical effects. This has been accomplished exemplary by addressing the TMOKE in transmission.

Numerical calculations showed that TMOKE for bare iron garnet film is very small $\delta \sim 10^{-5}$, both for transmitted and reflected light. When the film is covered by a smooth gold layer, the TMOKE is resonantly enhanced up to $\delta \sim 5 \cdot 10^{-3}$ but it can be observed only in reflection while the transmission almost vanishes. If the second feature – nanostructuring – comes into play extraordinary optical transmission appears with the giant TMOKE $\delta$ reaching $1.5 \cdot 10^{-2}$. A unique property of the TMOKE in transmission geometry is that the effect occurs solely due to the magnetic film boundary conditions and does not depend on the bulk component. This distinguishes the effect from all other magnetooptical effects in transmission and implies its potential use for applications in systems where subwavelength light localization is crucial.

Further, we demonstrate ultra high sensitivity and selectivity of the TMOKE in perforated systems. The high selectivity arises from the enhancement only for the resonances related to excitation of SPPs on the metal/magnetic dielectric interface. TMOKE is very sensitive to small deviations of light incidence from normal incidence. Moreover, it changes sign for SPPs traveling in opposite directions. Thus, TMOKE can become an important tool for complete characterization of plasmonic nanostructures.

Even though we paid here only attention to TMOKE, our previous study [30] allows us to claim that the proposed class of magnetoplasmonic heterostructures can be used to enhance also other magnetooptical effects such as the Faraday effect and the polar and the longitudinal Kerr effect. The appearance of these effects depends on the orientation of magnetization in the sample, which can be easily controlled by relatively small external magnetic field on the order of 100 Oe. Based on this performance these structures are very promising for applications in ultra high sensitivity devices and optical data processing. Moreover, relying on recent experimental results for the inverse Faraday effect [46] one may anticipate use of the proposed material for all-optical magnetic data storage devices operating at ultrafast (THz) frequencies.

**Supplementary: Methods**

**1. Calculation method**

Electromagnetic modeling of transmission and TMOKE spectra was performed on the basis of the rigorous coupled waves analysis (RCWA) technique [44] extended to the instance of gyrotropic materials [45]. Since the heterostructure is periodic, the electromagnetic field components in each layer can be represented as a superposition of Bloch waves. The Maxwell equations are written in a truncated Fourier space. The electromagnetic boundary conditions are then applied at the interfaces between the substrate region, the individual grating slabs, and finally the upper surface of the structure. The sequential application of electromagnetic boundary conditions reduces the computing effort for the reflected and the transmitted diffracted field amplitudes to the solution of a linear system of differential equations. To improve convergence of the method we employed the correct rules of Fourier factorization introduced in [45]. Using this method we obtained satisfactory convergence for 41 diffraction orders.

For the ferromagnetic film permittivity $\varepsilon_2$ and the absolute value of the gyration vector **g** we used our experimental data. Dispersion of both quantities was taken into account: e.g. at λ=650 nm $\varepsilon_2 = 5.12 + 0.018i$, $g = (3.4 + 0.4i) \cdot 10^{-3}$ and at λ=750 nm $\varepsilon_2 = 5.01 + 0.005i$, $g = (1.9 + 0.2i) \cdot 10^{-3}$. For the permittivity of gold we took the experimental data from [35].

## 2. Sample preparation

The magnetic layer of the magnetoplasmonic structure is a 2.5 micron thick bismuth-substituted rare-earth iron garnet film with composition $Bi_{0.4}(YGdSmCa)_{2.6}(FeGeSi)_5O_{12}$, grown by liquid phase epitaxy with a $Bi_2O_3 : PbO : B_2O_3$ melt on a gadolinium gallium garnet $Gd_3Ga_5O_{12}$ substrate with orientation (111). The film possesses an uniaxial magnetic anisotropy and a maze-like domain structure. The stripe domain width is 2.4 μm and the Curie temperature is 242 °C. The saturation magnetization of the film is $4\pi M_s$ = 453 G, the bubble collapse field is 238 Oe. The value of effective uniaxial magnetic anisotropy field $H_k^*$ = 1600 Oe, implying that application of an in-plane magnetic field with strength 1600 Oe leads to disappearance of the maze-like domain structure and results in a nearly complete in-plane magnetization orientation. The specific Faraday rotation is 0.46 deg/μm at wavelength 633 nm.

The magnetoplasmonic sample of structure shown in Fig.1 was fabricated by the following procedure. After initial cleaning of the iron garnet magnetic film in a $O_2$ plasma, a gold layer was deposited on it by a thermal evaporation process. PMMA 950 e-resist was spin coated on the gold layer at 3000 rpm and grating lines over a 1mm$^2$ area were drawn on the resist by electron beam lithography (Raith e-Line) using the Fixed Beam Moving Stage (FBMS) technique. For large area gratings, one has to take care of stitching errors, edge distortion, and proximity effect in addition to getting vertical walls of the grooves. We optimized the aperture, write field, dose and acceleration voltage to obtain the envisioned grating parameters. We used a 10 μm aperture with a write field of 100x100 μm$^2$, a dose of 110 μC/cm$^2$ and an acceleration voltage of 14 kV. The samples were then developed in Methyl Isobutyl Ketone (MIBK) for 90 sec at 21°C followed by rinsing in Isopropyl Alcohol (IPA) for 30 sec. In the next step, the pattern was transferred to the metal by reactive ion etching (RIE) using an Ar-ion plasma. The dry etch process was carried out to first calibrate the etch rates of both PMMA and Gold in $Cl_2$ and Ar plasmas. The Ar plasma process was found suitable to get no unpatterned Gold underneath the gratings and also to achieve the required grating depth. For achieving optimum grating parameters we have used thicker resist (300nm) and Gold layers (140nm). The etch process parameters were: etch time 7 min., flow rate 50 sccm, RF power 135W, and chamber pressure 0.2 Pa.

The sample was characterized by a AFM. The period and groove width were verified by SEM imaging. The grating parameters obtained are: depth of the grooves 120 nm, period 595 nm and air groove width 110nm.

## 3. Experimental set up and measurements

For magneto-optical measurements we used a halogen lamp as a source of white light. The collimated light was focused at the sample onto a spot with diameter of about 300-500 μm and an aperture angle below 1°. To perform measurements at different angles of light incidence the sample was mounted on a rotation stage. The zero-order transmission signal was spectrally dispersed with a single monochromator (linear dispersion 6.28 nm/mm) and detected with a charged coupled devices camera. The overall spectral resolution was below 0.3 nm. The polarization direction of the transmitted light was selected by a polarizer in the detection path. Magnetic fields up to 4 kOe were applied in transverse geometry using a water cooled electromagnet. During measurements the sample was at room temperature.

**References**

1. Heber, J. Plasmonics: Surfing the wave. *Nature* **461**, 720-722 (2009).
2. Polman, A. Plasmonics Applied. *Science* **322**, 868-869 (2008).
3. Maier, S. Special Issue: Plasmonics and Nanophotonics *Phys. Stat. Sol. RRL* **4**, A85–A98, 241–297 (2010).
4. Bozhevolnyi, S. I. *Plasmonics Nanoguides and Circuits.* (Pan Stanford Publ., 2008)
5. Najafov, H., Lee, B., Zhou, Q., Feldman, L. C. & Podzorov, V. Observation of long-range exciton diffusion in highly ordered organic semiconductors. *Nature Materials* **9**, 938-943 (2010).
6. Wurtz, G. A., Pollard, R. & Zayats, A. V. Optical Bistability in Nonlinear Surface-Plasmon Polaritonic Crystals. *Phys. Rev. Lett*. **97**, 057402-057406 (2006).
7. Kneipp, K. Surface-enhanced Raman Scattering. *Physics Today* 40-46 (2007).
8. Zvezdin, A. & Kotov, V. *Modern Magnetooptics and Magnetooptical Materials* (IOP, 1997).

9. Sarychev, A. K. & Shalaev, V. M. *Electrodynamics of Metamaterials.* (World Scientific, 2007).
10. Inoue, M., Arai, K., Fujii, T., & Abe, M. One-dimensional magnetophotonic crystals. *J. Appl. Phys.* **85**, 5768-5771 (1999).
11. Levy, M., Yang, H. C., Steel, M. J. & Fujita, J. Flat-Top Response in One-Dimensional Magnetic Photonic Bandgap Structures with Faraday Rotation Enhancement. *Lightwave Technol.* **19**, 1964-1970 (2001).
12. Zvezdin, A. K. & Belotelov, V. I. Magnetooptical properties of photonic crystals. *Euro. Phys. J. B* **37**, 479-487 (2004).
13. Ferguson, P. E., Stafsudd, O. M. & Wallis, R. F. Surface magnetoplasma waves in nickel. *Physica B & C* 86-88, 1403-1405 (1977).
14. Burke, J. J., Stegeman, G. I. & Tamir, T. Surface-polariton-like waves guided by thin, lossy metal films. *Phys. Rev. B* **33**, 5186-5201 (1986).
15. Hickernell, R. K. & Sarid, D. Long-range surface magnetoplasmons in thin nickel films. *Opt. Lett.* 12, 570-572 (1987).
16. Aers, G. C. & Boardman, A. D. The theory of semiconductor magnetoplasmon-polariton surface modes: Voigt geometry. *J. Phys. C: Solid State Phys.* **11**, 945-959 (1978).
17. Hickernell, R. K. & Sarid, D. Long-range surface magnetoplasmons in thin nickel films. *Opt. Lett.* **12**, 570-572 (1987).
18. Olney, R. D. & Romagnoli, R. J. Optical Effects of Surface Plasma Waves with Damping in Metallic Thin Films. *Appl. Opt.* **26**, 2279-2282 (1987).
19. Newman, D. M., Wears, M. L. & Matelon, R. J. Plasmon Transport Phenomena on a Continuous Ferromagnetic Surface. *Europhys. Lett.* **68**, 692-698 (2004).
20. Bonod, N., Reinisch, R., Popov, E. & Neviere, M. Optimization of surface-plasmon-enhanced magneto-optical effects. *J. Opt. Soc. Am. B* **21**, 791-797 (2004).
21. Gonzalez-Diaz, J. B., Garcia-Martin, A., Armelles, G., Garcia-Martin, J. M., Clavero, C., Cebollada, A., Lukaszew, R. A., Skuza, J. R., Kumah, D. P. & Clarke, R. Surface magneto-plasmon nonreciprocity effects in noble-metal/ferromagnetic heterostructures. *Phys. Rev. B* **76**, 153402 (2007).
22. Vila, E. F., Sueiro, X. M. B., Gonzalez-Diaz, J. B., Garcia-Martin, A., Garcia-Martin, J. M., Navarro, A. C., Reig, G. A., Rodriguez, D. M. & Sandoval, E. M. Surface Plasmon Resonance Effects in the Magneto-Optical Activity of Ag–Co–Ag Trilayers. *IEEE Trans. Magn.* **44**, 3303-3306 (2008).
23. Temnov, V., Armelles, G., Woggon, U., Guzatov, D., Cebollada, A., Garcia-Martin, A., Garcia-Martin, J. M., Thomay, T., Leitenstorfer, A. & Bratschitsch, R. Active magnetoplasmonics in hybrid metal/ferromagnet/metal microinterferometers. *Nature Photon.* **4**, 107 (2010).

24. Newman, D. M., Wears, M. L., Matelon, R. J. & Hooper, I. R. Magneto-optic behavior in the presence of surface plasmons. *J. Phys.: Condens. Matter* **20**, 345230(6p.) (2008).

25. Wurtz, G. A., Hendren W., Pollard, R., Atkinson, R., Le Guyader, L., Kirilyuk, A., Rasing, Th., Smolyaninov, I. I. & Zayats, A. V. Controlling optical transmission through magneto-plasmonic crystals with an external magnetic field. *New J. Phys.* **10**, 105012 (9p.) (2008).

26. Strelniker, Y. M. & Bergman, D. J. Transmittance and transparency of subwavelength perforated conducting films in the presence of a magnetic field. *Phys. Rev. B* **77**, 205113 (2008).

27. Clavero, C., Yang, K., Skuza, J. R. & Lukaszew, R. A. Magnetic-field modulation of surface plasmon polaritons on gratings. *Opt. Lett.* **35**, 1557-1559 (2010).

28. Armelles, G., Cebollada, A., Garcia-Martin, A., Garcia-Martin, J. M., Gonzalez, M. U., Gonzalez-Diaz, J. B., Ferreiro-Vila, E. & Torrado, J. F. Magnetoplasmonic nanostructures: systems supporting both plasmonic and magnetic properties. *J. Opt. A: Pure Appl. Opt.* **11**, 114023 (4p.) (2009).

29. Buchin, E. Y., Vaganova, E. I., Naumov, V. V., Paporkov, V. A. & Prokaznikov, A. V. Enhancement of the transversal magnetooptical Kerr effect in nanoperforated cobalt films. *Technical Phys. Lett.* **35**, 589-593 (2008).

30. V.I. Belotelov, L.L. Doskolovich, A.K. Zvezdin, Extraordinary magnetooptical effects and transmission through the metal-dielectric plasmonic systems, *Phys. Rev. Lett.* **98**, 77401(1-4) (2007).

31. V.I. Belotelov, D.A. Bykov, L.L. Doskolovich, A.N. Kalish, A.K. Zvezdin, "Extraordinary Transmission and Giant Magneto-optical Transverse Kerr Effect in Plasmonic Nanostructured Films", *Journal of the optical society of America B* **26**, 1594-1598 (2009).

32. Krinchik, G. S., Artem'ev, V.A. Magneto-optical properties of Ni, Co and Fe in ultraviolet visible and infrared parts of spectrum. *J. Exper. Theor. Phys.* **26**, 1080-1085 (1968).

33. Druzhinin, A. V., Lobov, I. D., Mayevskiy, V. M. & Bolotin, G. Transverse magnetooptical Kerr effect in transmission. *The Physics of Metals and Metallography* **56**, 58-65 (1983).

34. Landau, L. D., Lifshitz, E. M. *Electrodynamics of Continuous Media.* (Pergamon Press, 1984).

35. Johnson, P. B. & Christy, R. W. Optical Constants of the Noble Metals. *Phys. Rev. B* **6,** 4370-4376 (1972).

36. Dubovik, V. M. & Tosunyan, L. A. Toroidal moments in the physics of electromagnetic and weak interactions. *Sov. J. Part. Nucl.* **14**, 504-519 (1983).

37. Kalish, A. N., Belotelov, V. I. & Zvezdin, A. K. Optical properties of toroidal media. *SPIE - Int. Soc. Opt. Eng.* **6728**, 67283D (2007).
38. Ebbesen, T. W., Lezec, H. J., Ghaemi, H. F., Thio, T. & Wolff, P. A. Extraordinary optical transmission through sub-wavelength hole arrays. *Nature* **391**, 667-669 (1998).
39. Belotelov, V. I., Bykov, D. A., Doskolovich, L. L., Kalish, A. N., Zvezdin, A. K. Giant transversal Kerr effect in magnetoplasmonic heterostructures. *J. Exper. Theor. Phys.* **137**, 932-942 (2010).
40. Wood, R. W. Anomalous Diffraction Gratings. *Phys. Rev.* **48**, 928-936 (1935).
41. Sarrazin, M. & Vigneron, J. P. Bounded modes to the rescue of optical transmission. *Europhys. News* **38**, 27-31 (2007).
42. Porto, J. A., Garcia-Vidal, F. J. & Pendry, J. B. Transmission Resonances on Metallic Gratings with Very Narrow Slits. *Phys. Rev. Lett.* **83**, 2845-2848 (1999).
43. Marquier, F., Greffet, J., Collin, S., Pardo, F. & Pelouard, J. Resonant transmission through a metallic film due to coupled modes. *Opt. Express* **13**, 70-76 (2005).
44. Moharam, M. G., Pommet, D. A., Grann, E. B. & Gaylord, T. K. Stable implementation of the rigorous coupled-wave analysis for surface-relief gratings: enhanced transmittance matrix approachю *J. Opt. Soc. Am. A* **12**, 1077-1086 (1995).
45. Li, L. Fourier modal method for crossed anisotropic gratings with arbitrary permittivity and permeability tensors. *J. Opt. A: Pure Appl. Opt.* **5**, 345-355 (2003).
46. Kimel, A. V., Kirilyuk, A., Tsvetkov, A., Pisarev, R. V., Rasing, Th. Ultrafast non-thermal control of magnetization by instantaneous photomagnetic pulses. *Nature* **435**, 655-657 (2005).

**Acknowledgements**

This work was supported by Deutsche Forschungsgemeinschaft (DFG), the Russian Foundation for Basic Research (RFBR) and the Indian Department of Science and Technology (DST).

# Figures

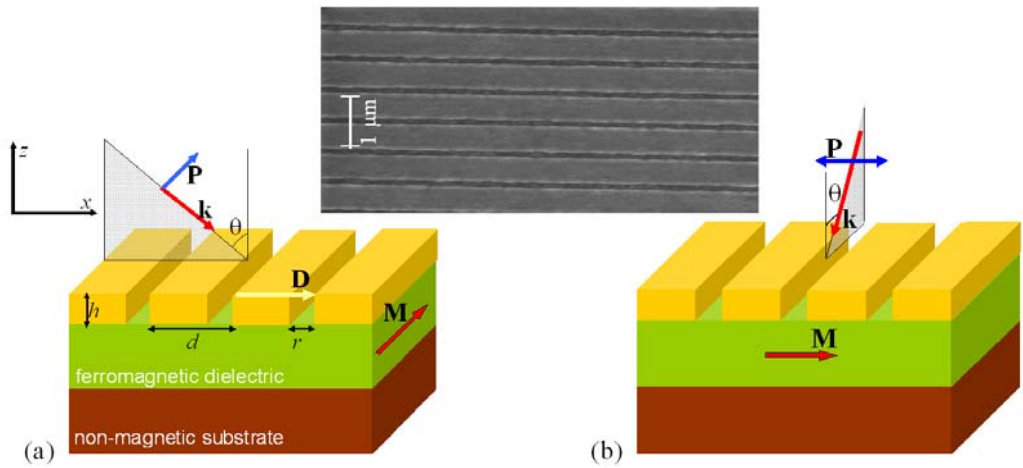

Fig. 1 Magnetoplasmonic heterostructure consisting of a periodically patterned gold layer on top of a bismuth iron garnet film grown on a gadolinium gallium garnet substrate. Two orientations of the magnetization **M** are considered: **M** is either parallel (a) or perpendicular (b) to the slits. The parameters of the perforated gold layer are $d$=594 nm, $r$= 110 nm, and $h$=120 nm. Inset: SEM image of the perforated gold structure.

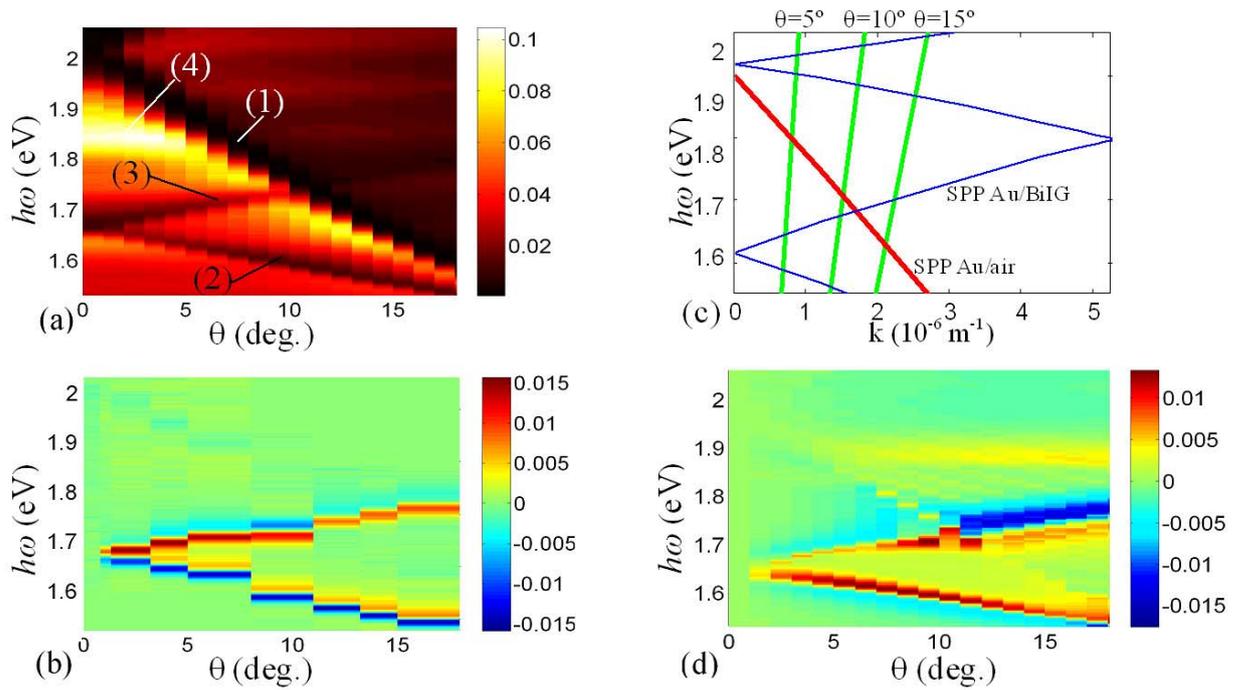

Fig. 2 Contour plots of measured and calculated transmission and TMOKE versus photon energy and angle of incidence (a-c) and dispersion diagram for SPPs (d) in the configuration with **M** along the slits and p-polarised incident light (see Fig. 1a). In-plane magnetic field is 2000 Oe. (a) experimentally measured transmission, (1) – (4) indicate transmission spectrum features related to SPPs or Fabry-Perot eigenmodes. (b) Experimentally measured relative change of transmission $\delta$. (c) Dispersion diagram for SPPs at the gold/air interface (thick red line) and at the gold/iron-garnet interface (thin blue lines), calculated in the free lattice approximation within the first Brillouin zone. Inclined green straight lines indicate the light lines for angles of incidence equal to 5°, 10° and 15°. (d) Calculated TMOKE parameter $\delta$.

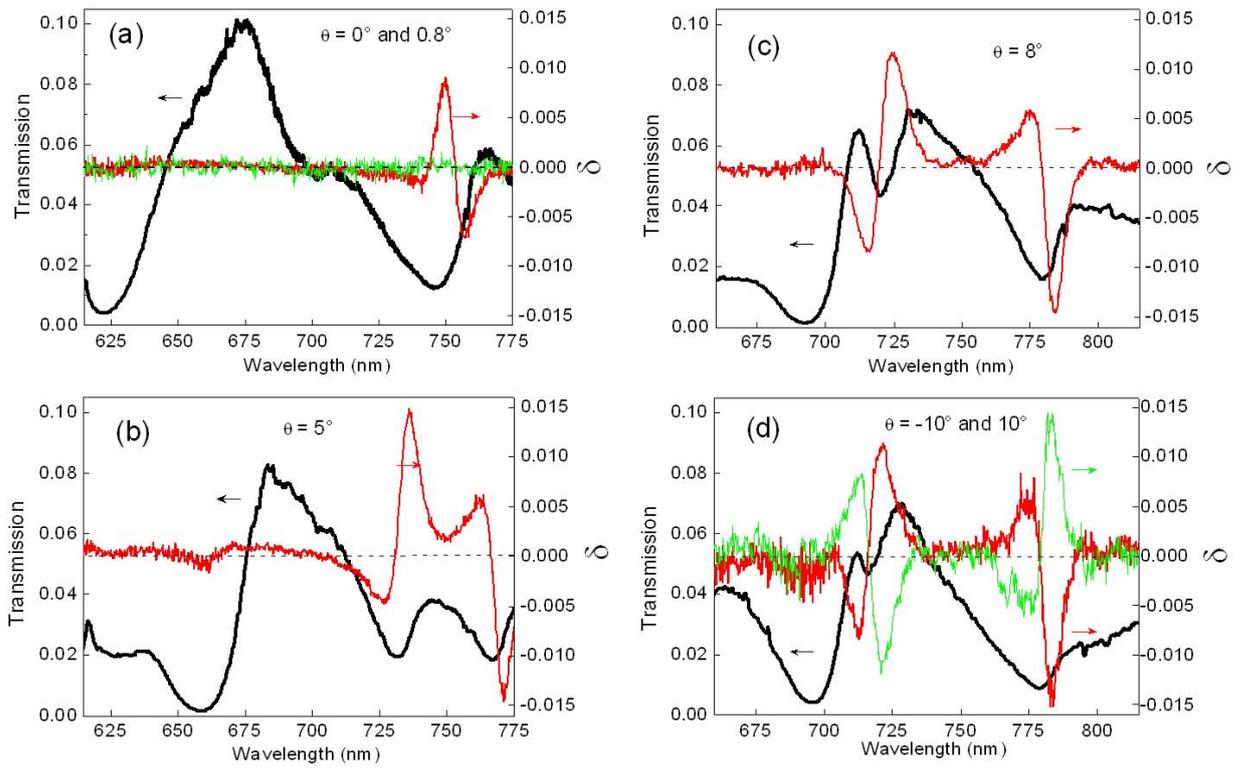

Fig. 3 Transmission (thick black curves) and TMOKE (thin red and green curves) spectra measured at different incidence angles: (a) θ=0º (green curve) and θ=0.8º (red curve); (b) θ=5º; (c) θ=8º; (d) θ=-10º (green curve) and θ=10º (red curve). In-plane magnetic field strength is 2000 Oe. **M** is along the slits (see Fig. 1a). Light is p-polarised.

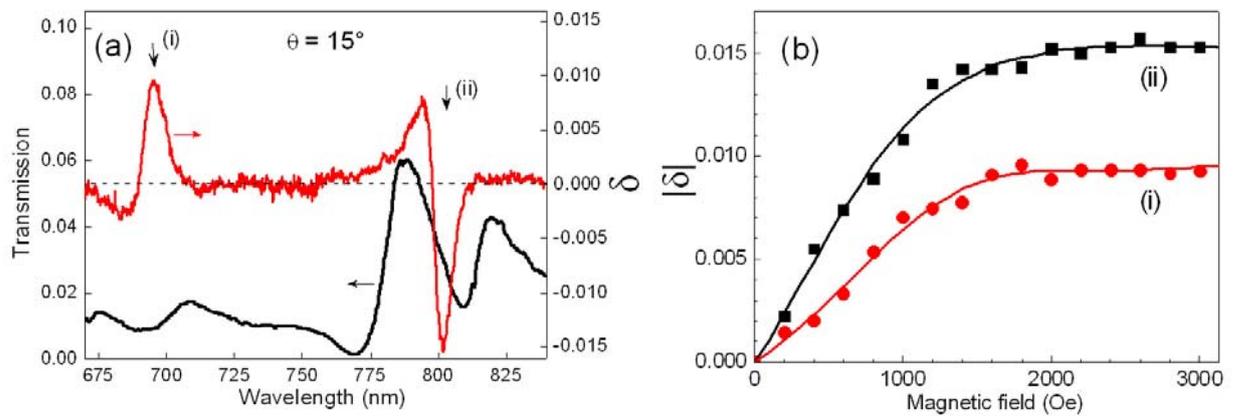

Fig. 4 Transmission and TMOKE spectra and magnetic field dependence of the TMOKE. (a) Transmission (thick black curve) and TMOKE (thin red curve) spectra measured at θ=15°. **M** is along the slits (see Fig. 1a), light is p-polarised. (b) Magnetic field dependence of the absolute value of δ at peaks (1) – black squares and (2) – red circles.

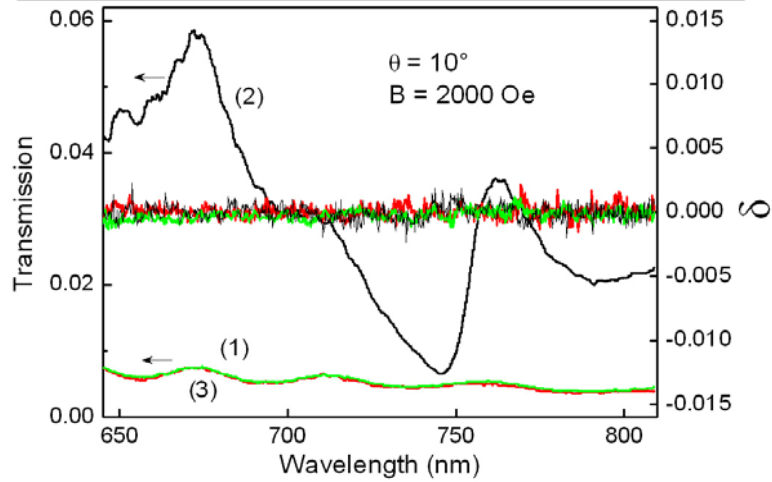

Fig. 5 Experimental transmission and TMOKE spectra for the configurations of (1) **M** parallel to the slits (the incidence plane is perpendicular to the slits), s-polarization of light (red curves), (2) **M** perpendicular to the slits (the incidence plane is along the slits), s-polarization (black curves), and (3) **M** perpendicular to the slits (incidence plane is along the slits), p-polarization (green curves). Table: different cases for the generation of SPPs and TMOKE.